\documentclass[12pt,reqno]{amsart}

\usepackage{amsmath,amssymb,amsfonts,latexsym}
\usepackage{graphicx,graphics}

\usepackage{mathrsfs}
\usepackage{fullpage}

\usepackage{graphicx}
\usepackage{epstopdf}
\newcommand{\ignore}[1]{}

\newcommand{\norm}[2]{\ensuremath{\left\|#1\right\|_{#2}}}

\newcommand{\mi}{\ensuremath{\mathrm{i}}}
\newcommand{\me}{\ensuremath{\mathrm{e}}}
\newcommand{\dif}{\ensuremath{\mathrm{d}}}
\newcommand{\pd}{\ensuremath{\partial}}

\newcommand{\CC}{\ensuremath{\mathbb{C}}}
\newcommand{\RR}{\ensuremath{\mathbb{R}}}
\newcommand{\NN}{\ensuremath{\mathbb{N}}}

\newcommand{\lam}{\ensuremath{\lambda}}
\newcommand{\eps}{\ensuremath{\epsilon}}

\newcommand{\equalsdef}{\stackrel{\mathrm{def}}{=:}}
\newcommand{\defequals}{\stackrel{\mathrm{def}}{:=}}
\renewcommand{\vec}[1]{\ensuremath{\mathbf{#1}}}
\newcommand{\mat}[1]{\ensuremath{\mathsf{#1}}}
\newcommand{\tr}{\ensuremath{\mathrm{t}}}
\newcommand{\beq}{\begin{equation}}
\newcommand{\eeq}{\end{equation}}
\DeclareMathOperator{\sech}{sech}

\DeclareMathOperator{\im}{Im}

\renewcommand{\Im}{\im}

\newcommand{\spm}{{\ensuremath{{\scriptscriptstyle\pm}}}}
\renewcommand{\sp}{{\ensuremath{{\scriptscriptstyle +}}}}
\newcommand{\sm}{{\ensuremath{{\scriptscriptstyle -}}}}

\numberwithin{equation}{section}

\newtheorem{theorem}{Theorem}
\newtheorem*{conjecture}{Conjecture}

\theoremstyle{definition}

\newtheorem{assume}[theorem]{Assumption}

\theoremstyle{remark}
\newtheorem{note}[theorem]{Note}
\newcommand{\br}{\begin{note}}
\newcommand{\er}{\end{note}}

\numberwithin{equation}{section}
\numberwithin{theorem}{section}

\title[The Zakharov--Shabat Problem]{The WKB approximation of semiclassical eigenvalues of the Zakharov--Shabat problem}%with smooth Klaus--Shaw potential}
\author{Yeongjoh Kim}
\address{Department of Mathematics, University of Wyoming, Laramie, WY 82071-3036, USA}
\email{ykim2@uwyo.edu}
\author{Long Lee}
\email{llee@uwyo.edu}
\author{Gregory D. Lyng}
\email{glyng@uwyo.edu}

\date{\today}
\begin{document}
\begin{abstract}
We numerically compute eigenvalues of the non-self-adjoint Zakharov--Shabat problem in the semiclassical regime. 
In particular, we compute the eigenvalues for a Gaussian potential and compare the results to the corresponding (formal) WKB approximations used in the approach to the semiclassical or zero-dispersion limit of the focusing nonlinear Schr\"odinger equation via semiclassical soliton ensembles. 
This numerical experiment, taken together with recent numerical experiments \cite{LLV,LL_PLA13}, speaks directly to the viability  of this approach; in particular, our experiment suggests a value for the rate of convergence of the WKB eigenvalues to the true eigenvalues in the semiclassical limit. This information provides some hint as to how these approximations might be rigorously incorporated into the asymptotic analysis of the singular limit for the associated nonlinear partial differential equation.  
\end{abstract}

\maketitle

\section{Introduction}\label{sec:intro}
\subsection{Eigenvalue problem, inverse spectral method}\label{ssec:evalpblm}
We consider the non-self-adjoint Zakharov--Shabat eigenvalue problem \cite{ZS}:
\begin{equation}\label{eq:zs}
\epsilon\frac{\dif}{\dif x}
\vec{w} =
\begin{bmatrix}
-\mi\lambda & \psi_0 \\
-\psi_0^* & \mi\lambda
\end{bmatrix}
\vec{w}\,.
\end{equation}
In equation \eqref{eq:zs}, we have written
\[
\vec{w}(x;\lam,\eps)=\begin{pmatrix} w_1(x;\lam,\eps) \\ w_2(x;\lam,\eps)\end{pmatrix}
\] 
for the solution, $\lambda\in\mathbb{C}$ is a spectral parameter, and the function $\psi_0:\RR\to\CC$ is a known potential. We suppose, to begin our discussion, that $\psi_0$ is specified by real-valued amplitude and phase functions $A_0$ and $S_0$, so that 
\[
\psi_0(x)=A_0(x)\exp\big(\mi S_0(x)/\eps\big)\,.
\]
We identify precise assumptions on $A_0$, $S_0$ in \S\ref{ssec:potentials} below. 
The parameter $\eps\in\RR$ is assumed to be positive but small; this introduces the ``semiclassical'' scaling. 
Our principal interest here is in those values of $\lambda\in\CC$ for which \eqref{eq:zs} has a solution in $L^2(\RR)^2$; these values comprise the discrete spectrum---the eigenvalues of \eqref{eq:zs}. 

Our motivation for analyzing \eqref{eq:zs} comes from its role in the theory of the initial-value problem for the cubic focusing nonlinear Schr\"odinger (NLS) equation 
\beq\label{eq:nls}
\mi\eps\frac{\pd\psi}{\pd t}+\frac{\eps^2}{2}\frac{\pd^2\psi}{\pd x^2}+|\psi|^2\psi=0\,.
\eeq
To emphasize the connection between \eqref{eq:zs} and \eqref{eq:nls}, we recall that Zakharov \& Shabat \cite{ZS} identified the linear spectral problem \eqref{eq:zs} as one half of the Lax pair for the NLS equation \eqref{eq:nls}. That is, the nonlinear equation \eqref{eq:nls} can be represented as the compatibility condition for two auxiliary linear problems---the Lax pair---and this structure allows one (in principle, at least) to construct solutions of the initial-value problem by the inverse spectral method (often called the inverse scattering transform). The initial step in this solution procedure is a spectral analysis of \eqref{eq:zs} in which the potential $\psi_0$ is taken to be the initial data for equation \eqref{eq:nls}, and the essential properties of the data for the initial-value problem are encoded in the spectral information---eigenvalues, norming constants, and reflection coefficient---associated with \eqref{eq:zs}. The temporal evolution is governed by properties of the other half of the Lax pair (details omitted here), is completely explicit, and takes place in the spectral domain. Finally, the solution at times $t>0$ is recovered by an inverse spectral transform; that is, the solution $\psi(x,t;\eps)$ is recovered from the time-evolved scattering data. A detailed discussion of this process for \eqref{eq:nls} can be found, for example, in the monographs \cite{APT,FT}.
 
\begin{note}[Semiclassical scaling]
We note that the small parameter $\eps$ appearing in \eqref{eq:nls} is the same as that appearing in the eigenvalue problem \eqref{eq:zs} above. In the NLS equation, the real parameter $\eps$ is a measurement of the ratio of dispersive effects to nonlinear ones. Our experiments here are focused on \eqref{eq:zs}, but they are motivated by a desire to understand the limiting behavior of solutions of the initial-value problem for \eqref{eq:nls} with fixed data in the singular limit $\eps\downarrow 0$. This zero-dispersion limit problem is sometimes called the semiclassical limit for the focusing NLS equation; the origin of this descriptor is based on the quantum-mechanical interpretation of the linear terms in \eqref{eq:nls}.
\end{note}

We recall that, in the inverse-spectral framework, the eigenvalues of \eqref{eq:zs} correspond to solitons, and these special solutions are fundamental elements of the theory of \eqref{eq:nls}. The remarkable properties of these solutions are well known; see, e.g., \cite{APT}.
%Indeed, the fact that a universal model equation like \eqref{eq:nls} supports soliton solutions might be regarded as one of the miracles of the theory of nonlinear waves. 
Thus, to summarize, given initial data for \eqref{eq:nls} or, equivalently, the potential in \eqref{eq:zs}, belonging to some reasonable class of functions (for example, $\mathscr{S}(\RR)$---the Schwartz class \cite{FT}), one would like to be able to effect a complete spectral analysis of \eqref{eq:zs} including, in particular, the location and multiplicity of the eigenvalues. Indeed, the eigenvalue locations have a direct impact on the dynamics and structure of the solution $\psi$ of \eqref{eq:nls}. Unfortunately, this is a challenge, and the general problem of rigorously extracting the requisite spectral information from \eqref{eq:zs} in the limit $\eps\downarrow0$  for a general potential $\psi_0$ remains largely open.   

\subsection{Known results}\label{ssec:known}
Despite the challenges that remain for the spectral analysis of \eqref{eq:zs} in general, there are a couple of important results that provide valuable guidance. Our discussion below assumes familiarity with the basic machinery and vocabulary of the inverse spectral method; we refer the interested reader who lacks this familiarity to the appendix of \cite{LLV} for a short but accessible outline of the steps in the inverse spectral method.

The first, most basic result of interest is due to  Satsuma \& Yajima \cite{SY}. They have shown that for a hyperbolic secant potential, i.e., 
\beq\label{eq:sydata}
\psi_0(x)=A\sech x\,,\quad A\in\RR\,,
\eeq
the eigenvalue problem (with $\eps=1)$ is \emph{exactly solvable}. In particular, Satsuma \& Yajima showed how to transform the equation \eqref{eq:zs} with potential given by \eqref{eq:sydata} to the hypergeometric equation which they were able to solve explicitly in terms of hypergeometric functions. In fact, they were able to write down formulae, in terms of the Gamma function, for the entries $a(\lambda)$ and $b(\lambda)$ in the scattering matrix,
\beq
\mat{S}(\lambda)=\begin{bmatrix} a(\lambda)^* & b(\lambda)^* \\ -b(\lambda) & a(\lambda) \end{bmatrix}\,,
\eeq
which relates the Jost solutions of \eqref{eq:zs} normalized at each of the spatial infinities. Importantly, these quantities give rise to the transmission and reflection coefficients, $T(\lambda)=1/a(\lambda)$ and $R(\lambda)=b(\lambda)/a(\lambda)$, that are essential ingredients in the solution of \eqref{eq:nls} by the inverse-spectral method. We recall that zeros of the analytic continuation of  the transmission coefficient $T$ to the upper half plane correspond to eigenvalues of \eqref{eq:zs}, and that $R$ is associated with continuous spectrum which, in this case, is confined to the real line. 

Inspecting Satsuma \& Yajima's formula, 
\beq
b(\lambda)=\frac{\mi |\Gamma(\mi\lambda+\frac{1}{2})|^2}{\Gamma(A)\Gamma(1-A)}=\mi\frac{\sin(\pi A)}{\cosh(\pi\lambda)}\,,
\eeq
we see that when $A=N\in\NN$, the reflection coefficient vanishes identically, and it turns out that the solution is a pure $N$-soliton, and the $N$ eigenvalues are also given explicitly; see \eqref{eq:syevals} below. Of particular interest is what happens when $N\to\infty$. As described by Lyng \& Miller \cite{LM} and in Note \ref{note:sy}, this problem is equivalent to a special case of the zero-dispersion limit problem for \eqref{eq:nls}.

\begin{note}[Non-zero phase]
Tovbis \& Venakides \cite{TV} have cleverly extended the above analysis to a one-parameter family of initial data of the form 
\[
\psi_\mathrm{tv}(x)=A_\mathrm{tv}(x)\exp(\mi S_\mathrm{tv}^\nu(x)/\eps)\,,
\]
where 
\[
A_\mathrm{tv}(x)=-\sech x\,,\quad \frac{\dif}{\dif x} S_\mathrm{tv}^\nu(x)=-\nu\tanh x\,.
\]
This is a particularly important result  as it provides a fundamental example in the case of nonzero phase. However, our focus here will be exclusively on real-valued, bell-shaped potentials like the hyperbolic secant considered by Satsuma \& Yajima.  Thus, for the remainder of the paper, we confine our attention to the case $S_0\equiv0$.
\end{note}

The second, more general, guiding result is more recent and is due to Klaus \& Shaw \cite{KS,KS2}. Their result says that, roughly speaking, eigenvalues for bell-shaped or ``single-lobe'' potentials are confined to the imaginary axis. Moreover, the eigenvalues are simple, and their number is given in terms of the $L^1$ norm of the potential. 
We give a precise statement of this result in Theorem \ref{thm:ks} below, and we use it to guide our numerical experiments. However, we note that it does not give detailed information about the precise locations of the eigenvalues.

\subsection{Semiclassical soliton ensembles}\label{ssec:sse}

We continue to focus on real-valued, bell shaped potentials. Given the dearth of detailed information about the eigenvalues of \eqref{eq:zs}, a standard procedure has been to replace the potential $\psi_0$ with an $\eps$-dependent reflectionless one, we shall denote it by $\psi_0^{(\eps)}$, whose eigenvalues are known exactly and are believed to be good approximations to the true (but unknown) eigenvalues corresponding to $\psi_0$; see Figure \ref{fig:cosine} (b) for an example of such a reflectionless potential. The mechanics of this process are described in more detail below in \S\ref{sec:sse}. Briefly, Ercolani et al.\ \cite{EJLM} have shown how to formally approximate the eigenvalue locations by exploiting a remarkable feature---known from the very beginning \cite{ZS}---of the problem \eqref{eq:zs}. Namely, in the small-$\eps$ limit, the nonselfadjoint problem \eqref{eq:zs} can be written as a semiclassical self-adjoint Schr\"odinger operator with a nonselfadjoint and formally small but $\lambda$-dependent correction. Ignoring this correction, one can apply standard results about the Schr\"odinger operator to obtain approximate eigenvalue locations which then satisfy a Bohr--Sommerfeld type quantization condition; see \eqref{eq:bs} below. 

These approximate eigenvalue locations were used in the monograph of Kamvissis et al.\ \cite{KMM} as the starting point for their asymptotic analysis. They neglected reflection,  and they and used the approximate WKB eigenvalues in place of the unknown true eigenvalues. This process creates a \emph{semiclassical soliton ensemble}---a sequence of exact multisoliton solutions of \eqref{eq:nls}. 

\begin{note}[The Satsuma--Yajima Ensemble]\label{note:sy}
In the special case that $\psi_0=A\sech x$, then with $\eps_N\defequals A/N$, this process reproduces the family of exact $N$-soliton solutions given by Satsuma \& Yajima, and the limit $N\to\infty$ is evidently a special case of the semiclassical limit, and there is no error induced by the use of $\psi_0^{(\eps)}$ in place of $\psi_0$. In general, however, this is not the case. This issue is addressed in \cite{LLV,LL_PLA13}. 
\end{note}

Our experiment here is part of a larger program to quantify the effects of this uncontrolled modification of the initial data for general bell-shaped potentials. 
We recall that the Whitham (or modulation) equations for \eqref{eq:nls} are \emph{elliptic}, and this feature of the problem confounds a common approach to similar problems which relies on local well-posedness of the hyperbolic Whitham system to permit an asymptotically vanishing perturbation of the eigenvalues. For example, Miller \cite{M} has rigorously shown that the WKB approximation at $t=0$ is asymptotically pointwise convergent, and more recent numerical experiments of Lee, Lyng, \& Vankova \cite{LLV} suggest convergence of the modified data to the true data in $L^2(\RR)$. However, it is not possible, on the basis of this information, to conclude that the solutions are close for any $t>0$. 
For further discussion of this point, see, e.g., \cite{KMM,M,LLV}

Remarkably, though, the numerical computations of Lee et al.\ \cite{LLV} suggest that this convergence indeed persists for $t>0$; we view this as quite intriguing, given the extreme modulational instabilities known to be present in the semiclassical regime. Indeed, in a follow-up work, Lee \& Lyng \cite{LL_PLA13} examined the sensitivity of of the semiclassical limit to qualitatively similar perturbations of the data, and they found that modulational instabilities almost instantly detected small, analytic perturbations of the data. These results give strong, but indirect, evidence that the WKB (approximate) eigenvalues used to generate the SSE are quite close to the true eigenvalues.

Here, continuing and complementing these investigations, we numerically measure the difference directly in the spectral plane. Indeed, we aim to quantify a rate of convergence of the approximate eigenvalues to the true eigenvalues as $\eps\downarrow 0$; our experiments suggest convergence at a rate of $O(\eps^2)$ as $\eps\downarrow0$. The experiment is described in \S\ref{sec:gsse}. We view this numerical experiment as a preliminary step toward incorporating the WKB approximation into a rigorous analysis built on the framework of Kamvissis et al.\ \cite{KMM}.  Assuming that the rate of convergence established here can be established rigorously (see the discussion in \S\ref{sec:discuss}), this would be a major step towards the development of a rigorous theory for the semiclassical limit of \eqref{eq:nls} that incorporates data beyond two special, exactly solvable, cases. Admittedly, the extension to analytic, bell-shaped, real data may seem at first glance to be a quite modest improvement, but this goal is at least a tractable target. The extension---mandated by the needs of applications---of the theory to more general (for example, non-analytic) data appears to be, for now, effectively out of reach.

\subsection{Plan}
In \S\ref{sec:frame} we begin by specifying the nature of the potentials $\psi_0$ that we will consider in \eqref{eq:zs}, and we describe a couple of important features of the eigenvalue problem. We give a careful statement of the spectral confinement results of Klaus \& Shaw \cite{KS,KS2}.
In \S\ref{sec:sse}, we outline the fundamental elements of the WKB approximation to the eigenvalues. In \S\ref{sec:nm}, we describe and validate the numerical method, and in \S\ref{sec:gsse}, we perform the main numerical experiment of the paper. As in previous work \cite{LLV,LL_PLA13}, our numerical experiment focuses on the case that $\psi_0(x)=\exp(-x^2)$. Finally, in \S\ref{sec:discuss}, we put the results in context and speculate about the implications of these calculations for the  zero-dispersion limit problem. 

\section{Framework: the Zakharov--Shabat eigenvalue problem}\label{sec:frame}
\subsection{Potentials}\label{ssec:potentials}
In this note, we restrict our attention to analytic Klaus--Shaw potentials . That is, we work in the framework of Kamvissis et al.\ \cite{KMM}, and we restrict our attention to potentials (initial data) of the form
\begin{equation}
\psi_0(x)=A_0(x)\,,
\label{eq:amplitude}
\end{equation}
where $A_0:\mathbb{R}\to(0,A]\subset\RR$ is even, bell-shaped, and real analytic. 
More precisely, $A_0$ is assumed to satisfy the assumptions detailed below.

\begin{assume}[Analytic Klaus--Shaw Potentials]\label{ass:pot}
The potential $A_0:\RR\to\RR$ is assumed to satisfy all of the following properties.
\begin{description}
\item[Decay] There exists $\alpha>0$ such that $|A_0(x)|=O(\me^{-\alpha|x|})$ as $x\to\pm\infty$.
\item[Evenness] $A_0$ is an even function, i.e., $A_0(x)=A_0(-x)$ for all $x\in\mathbb{R}$.
\item[Single Maximum]  $A_0$ has a single genuine maximum at $x=0$, i.e., $A_0(0)=A$, $A_0'(0)=0$, $A_0''(0)<0$.
\item[Analyticity] $A_0$ is real analytic. 
\end{description}
\end{assume}

\br
\noindent
\begin{enumerate}
\item[(a)] For the numerical calculations in this note, we shall restrict ourselves to the two concrete cases
\[
A_0(x)=A\sech x\,,\quad\text{and}\quad A_0(x)=\exp(-x^2)\,.
\]
Evidently, these choices fall into the category of Klaus--Shaw potentials described above. The sensitivity of the semiclassical limit problem for \eqref{eq:nls} to nonanalytic data has been investigated at the level of the partial differential equation by Clarke \& Miller \cite{CM}; at the spectral level, the sensitivity of the spectrum of \eqref{eq:zs} to nonanalytic perturbations of the potential was investigated by Bronski \cite{B}.  
\item[(b)] The spectral confinement result of Klaus \& Shaw (cf. Theorem \ref{thm:ks} below) does not require such stringent restrictions on the potential. For example, in \cite{KS}, Klaus \& Shaw assume---in addition to the essential single-lobe requirement---that $A_0\in L^1(\RR)$ is nonnegative, bounded, and piecewise smooth. However, analyticity is important for the analysis of \cite{KMM}; this is due to the ellipticity of the Whitham equations. There are questions about the ``stability'' of the limit, even within the analytic class \cite{CM,LL_PLA13}. The other, apparently unneeded, conditions (e.g., $A_0''(0)<0$) are used in the analysis of \cite{KMM} to guarantee that the WKB formulae below are sufficiently well behaved.
\end{enumerate}
\er

\subsection{About the eigenvalue problem: symmetry}\label{ssec:symmetry}
It is known that \eqref{eq:zs} is not self adjoint \cite{M_PD01}; thus, a priori, there is no restriction on where, in the complex plane, the spectrum may be. 
We observe that the eigenvalue problem \eqref{eq:zs} can be recast as
\beq\label{eq:zsd}
\mathscr{L}^{(\eps)}\vec{w}=\lambda\vec{w}\,,
\eeq
where $\mathscr{L}^{(\eps)}$ is the non-self-adjoint Dirac operator defined by 
\[
\mathscr{L}^{(\eps)}\defequals
\begin{bmatrix}
\mi\eps\frac{\dif}{\dif x} & -\mi A_0 \\
-\mi A_0 & -\mi\eps\frac{\dif}{\dif x}
\end{bmatrix}\,.
\]
It is a simple exercise to verify that if $\lambda$ is an eigenvalue of \eqref{eq:zs} with eigenfunction $\vec{w}=(w_1,w_2)^\tr$, then so is $\lambda^*$ with eigenfunction $\tilde{\vec{w}}=(w^*_2,- w^*_1)^\tr$. Thus, we will follow the established convention of considering and counting only eigenvalues $\lambda$ with $\Im\lambda>0$. Also, we note that if $A_0\in\mathscr{S}(\RR)$, then the $L^2(\RR)$ spectrum is comprised of the the continuous spectrum, which satisfies $\sigma_{\mathrm{cts}}(\mathscr{L}^{(\eps)})=\RR$, and a discrete set (possibly empty) of simple eigenvalues in the complex plane \cite{EJLM}.

\subsection{Klaus \& Shaw: spectral confinement}
In fact, for smooth Klaus--Shaw potentials, more is known. For each $\epsilon>0$ and for $\psi_0=A_0$ as described above, Klaus \& Shaw \cite{KS,KS2} have shown that the discrete spectrum of \eqref{eq:zs} is confined to the imaginary axis.
Indeed, for bell-shaped functions $A_0$ as described above, Klaus \& Shaw have shown the following.
\begin{theorem}[Klaus \& Shaw \cite{KS,KS2}]\label{thm:ks}
For $A_0$ as in Assumption~\ref{ass:pot}, eigenvalue problem \eqref{eq:zs} has precisely $N$ simple, purely imaginary eigenvalues with positive real parts where 
\beq\label{eq:n}
N=\left\lfloor \frac{1}{2}+\frac{1}{\eps\pi}\norm{A_0}{L^1(\RR)}\right\rfloor\,,
\eeq
and $\lfloor h\rfloor$ is the integer part of $h$. 
\end{theorem}
%\er

\section{Semiclassical soliton ensembles and the WKB approximation}\label{sec:sse}
\begin{figure}[ht] %  figure placement: here, top, bottom, or page
   \centering
   \includegraphics[width=3in]{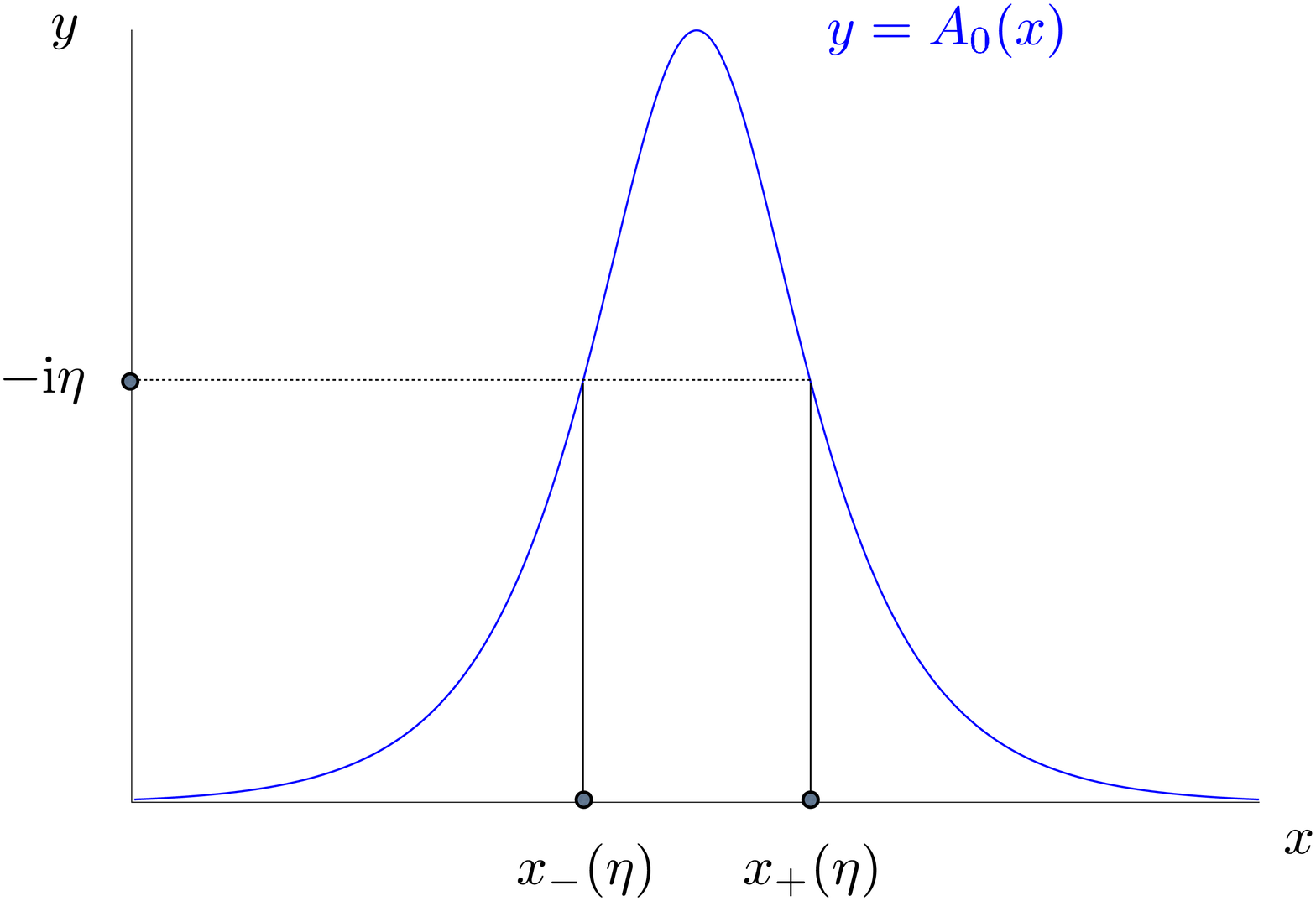} 
   \caption{The turning points $x_\pm(\eta)$.}
   \label{fig:tp}
\end{figure}

We begin by recalling the basic formulae for the WKB eigenvalues of \eqref{eq:zs}; for more details see \cite{EJLM,KMM,LLV}. We define the density function for $\eta\in(0,\mi A)$ via
\begin{equation}
\rho^0(\eta)\defequals\frac{\eta}{\pi}\int_{x_-(\eta)}^{x_+(\eta)}\frac{\dif x}{\sqrt{A_0(x)^2+\eta^2}}
=\frac{1}{\pi}\frac{\dif}{\dif\eta}\int_{x_-(\eta)}^{x_+(\eta)}\sqrt{A_0(x)^2+\eta^2}\,\dif x\,,
\label{eq:rho0}
\end{equation}
where $x_\pm(\eta)$ are the two real turning points; see Figure \ref{fig:tp}.
Using $\rho^0$ we next define 
\beq\label{eq:theta0}
\theta^0(\lambda)\defequals-\pi\int_\lambda^{\mi A}\rho^0(\eta)\,\dif\eta\,;
\eeq
the function $\theta^0$ gives a measure of the number of WKB eigenvalues on the imaginary axis between $\lam$ and $\mi A$.
To finish the process, we identify a sequence of distinguished values of $\eps$; we put
\beq
\epsilon_N\defequals-\frac{1}{N}\int_0^{\mi A}\rho^0(\eta)\,\dif\eta \label{eq:hbarn} 
%&\stackrel{\eqref{eq:rho0}}{=}-\frac{1}{N}\int_0^{\mi A}\frac{1}{\pi}\frac{\dif}{\dif\eta}\left[\int_{x_-(\eta)}^{x_+(\eta)}\sqrt{A_0(x)^2+\eta^2}\,\dif x\right]\,\dif\eta \nonumber \\
%&= -\frac{1}{\pi N}\left[\int_{x_-(\mi A)}^{x_+(\mi A)}\sqrt{A_0(x)^2-A^2}\,\dif x -\int_{x_-(0)}^{x_+(0)}\sqrt{A_0(x)^2}\,\dif x\right] \nonumber \\ 
=\frac{1}{\pi N}\int_{-\infty}^\infty A_0(x)\,\dif x\,,\,\quad N=1,2,3,\ldots.
\eeq
Finally, the WKB eigenvalues $\lambda^\mathrm{wkb}_{N,k}$ are defined (there are $N$ of them for each $\epsilon_N$) by the formula
\begin{align}
-\int_{\lambda^\mathrm{wkb}_{N,k}}^{\mi A}\rho^0(\eta)\,\dif \eta &=\epsilon_N\left(k+\frac{1}{2}\right)=\frac{\theta^0(\lambda^\mathrm{wkb}_{N,k})}{\pi}\,,\quad k=0,\ldots, N-1\,. \label{eq:bs}
\end{align} 
%Using the above formulae we may ``simplify'' the left-hand side:
%\begin{align*}
%-\int_{\lambda^{\mathrm{wkb}}_{N,k}}^{\mi A}\rho^0(\eta)\,\dif \eta&\stackrel{\eqref{eq:rho0}}{=}
%-\int_{\lambda^\mathrm{wkb}_{N,k}}^{\mi A}\frac{1}{\pi}\frac{\dif}{\dif\eta}\left[\int_{x_-(\eta)}^{x_+(\eta)}\sqrt{A_0(x)^2+\eta^2}\,\dif x\right]\,\dif\eta  \\
%&=-\frac{1}{\pi}\left[\int_{x_-(\mi A)}^{x_+(\mi A)}\sqrt{A_0(x)^2-A^2}\,\dif x -\int_{x_-(\lambda_{N,k})}^{x_+(\lambda_{N,k})}\sqrt{A_0(x)^2+\lambda_{N,k}^2}\,\dif x\right]  \\
%&=\frac{1}{\pi}\int_{x_-(\lambda_{N,k})}^{x_+(\lambda_{N,k})}\sqrt{A_0(x)^2+\lambda_{N,k}^2}\,\dif x \\
%&=\frac{2}{\pi}\int_{0}^{x_+(\lambda^\mathrm{wkb}_{N,k})}\sqrt{A_0(x)^2+(\lambda^\mathrm{wkb}_{N,k})^2}\,\dif x\,.
%\end{align*}
%where the last equality follows from our assumption that $A_0$ is even. 
Therefore, writing $\lambda^\mathrm{wkb}_{N,k}=\mi \tau^\mathrm{wkb}_{N,k}$ for $\tau^\mathrm{wkb}_{N,k}\in(0,A)\subset\mathbb{R}$, we obtain the WKB eigenvalues as solutions to the equations 
\begin{equation}
\int_0^{x_+(\mi t_{N,k})}\sqrt{A_0(x)^2-(\tau_{N,k}^{\mathrm{wkb}})^2}\,\dif x=\frac{\pi\epsilon_N}{2}\left(k+\frac{1}{2}\right)\,,
%\;\;k=0,1,2,\ldots N-1\,.
\label{eq:wkbeval}
\end{equation}
for $k=0,1,\ldots,N-1$.
%In this case, the auxiliary scattering data (proportionality constants) are given by 
%\beq\label{eq:gammak}
%\tilde\gamma_{N,k}=(-1)^{k+1}\,.
%\eeq
Specializing the above discussion to the case that the potential $A_0$ is given by 
\begin{equation}
\psi_0(t)=A_0(x)=\me^{-x^2}\,,
\label{eq:gauss}
\end{equation}
we find, from \eqref{eq:hbarn},
\begin{equation}\label{eq:epsilon_N}
\epsilon_N=\frac{1}{\pi N}\int_{-\infty}^\infty\me^{-t^2}\,\dif t =\frac{1}{\sqrt{\pi}N}\,,
\end{equation}
and, from formula \eqref{eq:wkbeval},
\begin{equation}\label{eq:gausswkb}
\int_0^{x_\sp(\mi \tau_{N,k}^\mathrm{wkb})}\sqrt{\me^{-2x^2}-(\tau_{N,k}^\mathrm{wkb})^2}\,\dif x=\frac{\sqrt{\pi}}{2N}\left(k+\frac{1}{2}\right),\quad k=0,1,2,\ldots N-1\,.
\end{equation}
In this case the turning points $x_\spm$ are given explicitly by 
\begin{equation}\label{eq:gxplus}
x_\spm(\mi \tau)=\pm\sqrt{-\ln \tau}\,,
\end{equation}
and equation \eqref{eq:wkbeval} can be rewritten as  
\begin{equation}
\int_0^{\sqrt{-\ln \tau_{N,k}^\mathrm{wkb}}}\sqrt{\me^{-2x^2}-(\tau_{N,k}^\mathrm{wkb})^2}\,\dif x=\frac{\sqrt{\pi}}{2 N}\left(k+\frac{1}{2}\right)\,,\quad
k=0,1,2,\ldots, N-1\,.
\label{eq:gwkbeval}
\end{equation}
This equation was solved to high precision by Lee et al.\ \cite{LLV}, and the 250-digit accuracy of the obtained solutions was verified using both \textsc{Mathematica} and \textsc{Maple} routines. We report the computed values in Appendix \ref{sec:wkbevals} below. Numerical experiments \cite{LLV,LL_PLA13} suggest, indirectly, that these values are quite close to the \emph{true} eigenvalues of \eqref{eq:zs} and that the distinct eigenvalues coalesce in the limit $\eps\downarrow 0$. In the next section we describe a numerical method aimed at accurately approximating the differences for a number of values of $\eps$ so that proximity of the WKB eigenvalues to the true eigenvalues can be measured directly and a rate of convergence, as $\eps\downarrow0$, can be estimated. 

\section{Asymptotic analysis \& numerical method}\label{sec:nm}
\subsection{Evans-function analysis}\label{ssec:evans}
We shall numerically compute eigenvalues of the Zakharov--Shabat problem \eqref{eq:zs} by means of a complex shooting method originally proposed and implemented by Bronski \cite{B}. Here, we outline how this method approximates zeros of the Evans function (or transmission coefficient) associated with \eqref{eq:zs}. This connection may be useful; in \S\ref{sec:discuss} we describe some possible future projects related to \eqref{eq:zs} that exploit recent developments in methods for the numerical approximation of Evans functions (see, e.g., \cite{HLZ_ARMA09,HL,HSZ,HZ}).

First, we note that the eigenvalue problem \eqref{eq:zs} 
can be reformulated as
\beq\label{eq:zs2}
\vec{w}'=\mat{A}(x;\lam,\eps)\vec{w}\,,
\eeq
where 
\begin{align}
\mat{A}(x;\lam,\eps)& \defequals
\begin{bmatrix}
-\mi\lam/\eps & A_0(x)/\eps \\
-A_0(x)/\eps & \mi\lam/\eps
\end{bmatrix} \\
	&=\begin{bmatrix}
-\mi\lam/\eps & 0 \\
0 & \mi\lam/\eps
\end{bmatrix}+
\begin{bmatrix}
0 & A_0(x)/\eps \\
-A_0(x)/\eps & 0
\end{bmatrix}\, \nonumber\\
	&\equalsdef\mat{A}_\infty(\lam,\eps)+\mat{B}(x;\eps)\,.
\end{align}
We observe that, due to the Assumption \ref{ass:pot} on the potential $A_0$, we find that there exists $\alpha>0$ such that 
\beq\label{eq:expdecay}
\norm{\mat{A}(x;\lam,\eps)-\mat{A}_\infty(\lam;\eps)}{}=\norm{\mat{B}(x;\eps)}{}=O(\me^{-\alpha|x|})\quad
\text{as}\;\;x\to\pm\infty\,.
\eeq
Here, $\|\cdot\|$ is any matrix norm. 
Evidently, the eigenvalues of the limiting matrix $\mat{A}_\infty$ are given by 
\beq\label{eq:mu}
\mu_\sm(\lam,\eps)=-\frac{\mi\lam}{\eps}\,,\quad
\mu_\sp(\lam,\eps)=+\frac{\mi\lam}{\eps}\,,
\eeq
and the corresponding eigenvectors are 
\(
\vec{v}_\sm=\vec{e}_1
\)
and 
\(
\vec{v}_\sp=\vec{e}_2
\).
Under the assumption \eqref{eq:expdecay}, for fixed $\eps>0$ and for $\lambda\in\{z\in\CC\,:\,\im z>0\}$, we find that there are solutions 
\[
\vec{w}_\spm(x;\lam,\eps)
\]
of \eqref{eq:zs2} which approach the decaying solutions $\vec{y}_\spm(x)=\exp(\mu_\spm(\lambda,\eps))\vec{v}_\spm$ of the limiting constant-coefficient system 
\[
\vec{y}'=\mat{A}_\infty(\lambda,\eps)\vec{y}
\]
as $x\to\pm\infty$. Then, up to a non-vanishing analytic factor, the Evans function, an analytic function on its natural domain, is given by
\beq\label{eq:evans}
D^{(\eps)}(\lam)\defequals\det\big(\vec{w}_\sp(x;\lam,\eps),\vec{w}_\sm(x;\lam,\eps)\big)|_{x=0}\,.
\eeq
An immediate consequence of this definition is that $D^{(\eps)}(\lambda)=0$ if and only if $\lambda$ is an eigenvalue. For, $D^{(\eps)}$ detects a linear dependence between solutions of \eqref{eq:zs2} decaying at $\pm\infty$. As described below, Bronski's shooting method is based on approximating $\vec{w}_\sm$ and determining the values of $\lambda$ for which such linear dependence exists. For this determination, analyticity is used in an essential way.

\subsection{Numerical Methods}\label{sec:nm-Bronski}

As noted above, we adopt the shooting method of Bronski \cite{B96} to locate the eigenvalues for \eqref{eq:zs}. However, to make the discussion here self-contained, we give a thorough description of the procedure. 
%Furthermore, in each step of the procedure, we also report the numerical difficulties, to our knowledge, and suggest treatments to overcome these difficulties. 

\begin{description}
\item[Step\#1 (Spatial Integration)] Conventionally, the process of solving the eigenvalue problem begins with integrating the differential equation \eqref{eq:zs} for fixed $\lambda$ with the initial conditions
\beq
\vec{w}(-L) = \begin{pmatrix} 
1 \\
0 
\end{pmatrix}
=\vec{v}_\sm
\eeq
to $x=+L$, where $L$ is chosen so that $\psi_{0}(\pm L)\approx 0$. However, direct numerical integration of this system suffers from large numerical errors due to the exponential growth of the mode corresponding to $\mu_\sm(\lam,\eps)$ at large $L$ when $\im\lambda$  is large and $\eps$ is small. To eliminate this growth, for our numerical calculations, we define
\beq
\vec{w}(x)=\me^{\mu_\sm(\lam,\eps)x}\vec{u}(x)
\eeq
and we integrate 
\beq
\frac{\dif \vec{u}}{\dif x}=(\mat{A}-\mu_\sm\mat{I})\vec{u}
\eeq
from $x=-L$ to $x=+L$. The specified data at $x=-L$ is given by 
\[
\vec{u}(-L)=\begin{pmatrix} \exp\big(-\mu_\sm(\lambda,\eps)L\big) \\ 0 \end{pmatrix}
\]
corresponding to the choice $\vec{w}(-L)=(1,0)^\tr$.  We use a 6th-order Runge--Kutta scheme developed in \cite{Butcher64} as the integrator; we typically take $L=40$ and $\Delta x=0.002$. 
\br[Quantitative Gap Lemma]
Using the known decay rate of $A_0$, it is possible to quantify the size of the initialization error that arises from truncating the domain of \eqref{eq:zs} and integrating from $x=-L$. For example, this kind of analysis has been done by Humpherys et al.\ \cite{HLZ_ARMA09}---using the ``quantitative gap lemma''---in their numerical approximation of the Evans function associated with viscous shock-layer solutions of the compressible Navier--Stokes equations. On the other hand, the true error is based on the values computed at $x=+L$, and our validation process (see \S\ref{ssec:validate} below) suggests that these errors are quite small. We therefore omit a detailed analysis of the initialization error.
%Q? What about the flow to $L=+40$?
\er

\item[Step \#2 (Integrand Assembly)]
We write a generic complex number $\lam$ in terms of its real and imaginary part as $\lam=\gamma+\mi\tau$, and we suppose that 
\beq
\Gamma^{j}:[0,1]\to\CC\,,\quad j=1,2,3,4\,;
\eeq
are the four sides of a rectangle in the complex plane. Here, $(j=1)\equiv\text{top}$,  $(j=2)\equiv\text{right side}$,  $(j=3)\equiv\text{bottom}$, and  $(j=4)\equiv\text{left side}$.  We adopt the following labeling convention for the grid points on $\Gamma^{j}$:
\beq\label{eq:lamb-nj}
\lambda_{n}^{j}=\gamma_n^{j}+\mi\tau_n^{j}\,,\quad n=1,\ldots,M^j\,.
\eeq
Evidently, the corner points carry multiple labels. That is, for example, 
\[
\lambda_1^{1}=\lambda_M^{4}
\]
and so on.  At this point, we use the labels 
\beq
\begin{split}
\vec{w} = \begin{pmatrix}  w_{1} \\ w_{2} \end{pmatrix},\quad
\vec{u} = \begin{pmatrix}  u_{1} \\ u_{2} \end{pmatrix},
\end{split}
\eeq
and we are interested in finding zeros of $w_{1}(L;\lambda,\eps)$. To this end,  we compute for the selected $\lam$-values the quantity 
\beq\label{eq:integrand}
f(\lam_n^{j},\eps)=\frac{u_1'(L;\lam_n^{j},\eps)}{u_1(L;\lam_n^{j},\eps)}\,.
\eeq
For notational convenience, we denote the numerator and denominator of \eqref{eq:integrand} by $u_1'(\lam_n^j)$ and $u_1(\lam_n^j)$ respectively.
To evaluate \eqref{eq:integrand}, we need to approximate the derivative with respect to $\lambda$ in the numerator. Instead of using the finite difference approximation, to obtain a third-order approximation of the derivative $u_1'(\lam_n^{j})$, after computing $u_1(\lam_n^{j})$, we use the cubic spline (with not-a-knot endpoint condition) to interpolate $u_1(\lam_n^{j})$ at $\lambda_{n}^{j}$, knowing that the coefficient of the spline function gives the derivative $u_1'(\lam_n^{j})$ at $\lambda_{n}^{j}$. 

%Here, 
%\[
%\Delta \lambda^j=
%\begin{cases}
% & j=1,3 \\
% & j=2,4
%\end{cases}
%\]
%
%\vskip 12pt
%
\item[Step \#3 (Moment Calculations)]
Suppose that $\Omega\subset\CC$ is a simply connected domain. If $\Gamma$ is a simple closed curve in $\Omega$ and if $f$ is holomorphic on $\Omega$ with zeros 
\(
\lam_1\,,\ldots,\lam_N
\)
inside $\Gamma$, then the $p$th moment of $f$ about $z_0$ is given by 
\beq\label{eq:moment}
M_p(z_0)=\frac{1}{2\pi\mi}\oint_\Gamma\frac{(\zeta-z_0)^pf'(\zeta)}{f(\zeta)}\,\dif \zeta\,,
\eeq
and 
\[
M_p(z_0)=\sum_{k=1}^N(\lam_k-z_0)^p\,.
\]
Thus, $M_0(0)$ returns the number of zeros inside the contour $\Gamma$, and $M_1(0)=\lam_1+\cdots+\lam_N$.
Therefore, provided that there is only one zero inside $\Gamma$, the first moment about zero returns its location. We may thus find eigenvalue locations by approximating integrals of the form \eqref{eq:moment}.
In our first numerical calculation, for the case $A_{0}(x)=A\sech(x)$---for which the eigenvalues are already known, we take $\Gamma$ to be a rectangle placed to enclose a solitary eigenvalue.  For the principal experiment, we center each of the rectangles $\Gamma_k$ at the approximate eigenvalue location given by the solution of \eqref{eq:gwkbeval}.  Now suppose that the four corners of a rectangular contour $\Gamma_{k}$ in the complex plane are labeled as shown in Figure \ref{fig:gamma}. 

\begin{figure}[ht] %  figure placement: here, top, bottom, or page
   \centering
   \includegraphics[width=3.5cm]{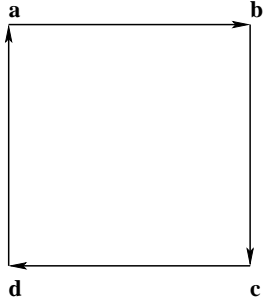} 
   \caption{The rectangular contour $\Gamma_{k}$.}
   \label{fig:gamma}
\end{figure}

Thus, using the definition (\ref{eq:lamb-nj}) and the integrand (\ref{eq:integrand}), the two moments along the rectangular contour $\Gamma_{k}$ can be expressed as 
\begin{subequations}\label{eq:moments}
\begin{multline}
n_{k}=\frac{1}{2\pi\mi}\left[\int_{a}^{b}\frac{u_1'(\lam^{1})}{u_1(\lam^{1})}\,\dif\gamma^{1}
+\mi\int_{b}^{c}\frac{u_1'(\lam^{2})}{u_1(\lam^{2})}\,\dif\tau^{2} \right.
\left.+\int_{c}^{d}\frac{u_1'(\lam^{3})}{u_1(\lam^{3})}\,\dif\gamma^{3} 
+\mi\int_{d}^{a}\frac{u_1'(\lam^{4})}{u_1(\lam^{4})}\,\dif\tau^{4} \right]\,,
\end{multline}
and
\begin{multline}
\ell_{k}=\frac{1}{2\pi\mi}\left[\int_{a}^{b}\frac{\lam^{1} u_1'(\lam^{1})}{u_1(\lam^{1})}d\gamma^{1} +\mi\int_{b}^{c}\frac{\lam^{2}u_1'(\lam^{2})}{u_1(\lam^{2})}d\tau^{2}\right.
 \\
\left.+ \int_{c}^{d}\frac{\lam^{3}u_1'(\lam^{3})}{u_1(\lam^{3})}d\gamma^{3} + \mi\int_{d}^{a}\frac{\lam^{4}u_{1}'(\lam^{4})}{u_1(\lam^{4})}d\tau^{4} \right].
\end{multline}
\end{subequations}
As described above, when $n_{k}\approx1$, the corresponding value of $\ell_{k}$ gives the approximate location of the eigenvalue enclosed by the contour $\Gamma_{k}$. For our numerical calculation, each integral in \eqref{eq:moments} is evaluated by the 6th-order Newton-Cotes integration formula (also referred to as Weddle's rule) \cite{DA08}.
\begin{note}
The superscripts in \eqref{eq:moments} refer to the labeling scheme described in \eqref{eq:lamb-nj} and should not be confused with exponents appearing in the moment formula \eqref{eq:moment}. Thus, $n_k$ is the zeroth moment about zero and $\ell_k$ is the first moment about zero.
\end{note}
\end{description}

\subsection{Validation: the Satsuma--Yajima ensemble}\label{ssec:validate}
To test the methodology, we look at the case 
\[
A_0(x)=A\sech x\,,
\]
for which the eigenvalues are known exactly \cite{SY}.
Indeed, following the notation of \cite{LM}, the $N$ eigenvalues of \eqref{eq:zs} are given by 
\beq\label{eq:syevals}
\lam_{N,k}^{\mathrm{sy}}\defequals\mi A-\mi\left(k+\frac{1}{2}\right)\eps_N\,,
\quad k=0,1,\ldots, N-1\,,
\eeq
and we recall that, in this case, 
\[
\eps_N\defequals A/N\,.
\]
We are thus considering a ``quantized'' sequence of values of $\eps$ which tends to zero as $N\to\infty$.

Following the algorithm described in \S\ref{sec:nm-Bronski}, we compute the eigenvalues for the cases $N=5, 10, 15$, and $20$. The $k^{th}$ eigenvalue approximation for the case of $N$ is denoted by $\lambda_{N,k}^\mathrm{app}=\mi\tau_{N,k}^\mathrm{app}$. We define the maximum error for each $N$ to be
\beq
e_{N}\defequals  \max_{k} | \lambda_{N,k}^\mathrm{app} - \lam_{N,k}^{\mathrm{sy}}|.
\eeq

\begin{table}[ht]
\begin{centering}
\caption{Validation: the Satsuma--Yajima ensemble}
\label{tab:syvalidate}
\begin{tabular}{c|ccccccc}
%$N$ & 5 & 10 & 15 & 20 \\ \hline
$N$ & $e_{N}$ & at $\lambda_{N, k}$ & $dx$ & $d\gamma$ & $d\tau$ & $\overline{ab}$ & $\overline{bc}$\\ \hline
5&6.7174E-10&$\lambda_{5,4}$&80/40,000&0.2/192&0.4/192&0.2&0.4\\ \hline
10&2.2438E-09&$\lambda_{10, 9}$&80/40,000&0.2/192&0.2/192&0.2&0.2\\ \hline
15&9.3286E-09&$\lambda_{15, 14}$&80/40,000&0.2/192&0.133/192&0.2&0.133\\ \hline
20&3.0100E-08&$\lambda_{20, 19}$&80/40,000&0.2/192&0.1/192&0.2&0.1\\ 
%$\max_{k}|\lambda_{N,k}^\mathrm{SY}-\lambda_{N,k}^\mathrm{app}|$ & TODO & 2.24E-09 & TODO & TODO \\
%$\eps_N^2$ & 1.6E-1 & 4E-2 & 1.78E-2 & 1.0E-2
\end{tabular}
\end{centering}
%\caption{Validation: the Satsumy--Yajima ensemble}
\end{table}

In Table \ref{tab:syvalidate} we list $e_{N}$ and the location at which the maximum error occurs. We also list the mesh size $dx$ used for the eigenvalue problem, and the mesh size $d\gamma$ and $d\tau$ used for the moment calculations. We discover that the largest errors always occur at $(N-1)^{th}$ eigenvalue, closest to the real-axis. This agrees with Bronski's finding \cite{B96} that the numerical method suffers near $\sigma_{\mathrm{cts}}(\mathscr{L}^{(\eps)})=\RR$. Indeed, as noted by Bronski,  the boundary conditions reverse roles in the lower half plane, and the method is not expected to be reliable close to the real line. We also see that the error increases when $\eps$ decreases. For the case of the smallest $\eps$ ($N=20$), we were able to control the error to the order of $10^{-8}$.

\section{The Gaussian Case}\label{sec:gsse}
\subsection{Experiment}
We now present the results of the principal calculation of the paper. We consider the Zakharov--Shabat problem \eqref{eq:zs} with Gaussian potential
\beq
\psi_0(x)=\exp(-x^2)\,.
\eeq 
In Step \#1 of the algorithm, the domain of calculation is $-40\le x\le 40$ ($L=40$), and the step size of the integration is $dx=80/40,000$. To test whether our computational results are numerically convergent, we use a finer mesh size $dx=80/80,000$ to compute the eigenvalue problem for the case of $N=15$.  We find that the difference between computed  eigenvalue location is of the order of $10^{-11}$ between the two different mesh sizes. In Step \#3, with the SSE eigenvalue located at the center of the rectangle, the length of the top and the bottom side of each rectangle is $\overline{ab} =\overline{cd}= 0.2$, while the left and the right side of the rectangle is $\overline{bc} =\overline{da}= 0.0815$. A total number of $193$ grid points are evaluated at each side of the rectangle ($M=193$ in equation \eqref{eq:lamb-nj}). This gives that $d\gamma \approx \Delta\gamma= 0.2/192$ and $d\tau \approx\Delta\tau=0.0815/192$ in Eq. (\ref{eq:moments}).  
We denote the difference (in absolute value) between $\lambda_{N,k}^\mathrm{app}$ and the $k^{th}$ WKB eigenvalue $\lambda_{N,k}^\mathrm{wkb}$ by
\beq
D_{k}^{N} =|\lambda_{N,k}^\mathrm{app}-\lambda_{N,k}^\mathrm{wkb}|, 
\eeq
where $\lambda_{N,k}^\mathrm{wkb}=\mi\tau_{N,k}^\mathrm{wkb}$ are the WKB eigenvalues computed in \cite{LLV}.  The computed values for $\tau_{N,k}^\mathrm{wkb}$ and $\tau_{N,k}^\mathrm{app}$ for $N=10,\ldots, 22$ are recorded below  in Appendix \ref{sec:wkbevals}.

%computed values for $\tau_{N,k}^\mathrm{wkb}$ for $N=5,10,15,20$ are recorded in Appendix \ref{sec:wkbevals}
%\begin{subequations}
%\begin{align}
%\Delta^{(N)}_M&\defequals\max_k|\lambda_{N,k}^\mathrm{app}-\lambda_{N,k}^\mathrm{wkb}|\,, \\
%\Delta^{(N)}_m&\defequals\min_k|\lambda_{N,k}^\mathrm{app}-\lambda_{N,k}^\mathrm{wkb}|\,, \\
%\intertext{and}
%\Delta^{(N)}_\mathrm{ave}&\defequals \frac{1}{N}\sum_{k=0}^{N-1}|\lambda_{N,k}^\mathrm{app}-\lambda_{N,k}^\mathrm{wkb}|\,.
%\end{align}
%\end{subequations}
%where $\lambda_{N,k}^\mathrm{wkb}=\mi\tau_{N,k}^\mathrm{wkb}$ are the WKB eigenvalues computed in \cite{LLV}; computed values for $\tau_{N,k}^\mathrm{wkb}$ for $N=5,10,15,20$ are recorded in Appendix \ref{sec:wkbevals}.
%\begin{table}[ht]
%\begin{centering}
%\begin{tabular}{c|ccccccccccc}
%$N$ & 10 & 11 & 12 & 13 & 14 & 15 & 16 & 17 & 18 & 19 & 20   \\ \hline
%$\Delta_m^{(N)}$ & 0.2066E-3 & 0.1702E-3 & 0.1426E-3 & 0.1212E-3
%&  0.1043E-3 & 0.0907E-3 & 0.0796E-3 &0.0705E-3 & 0.0629E-3 & 0.0564E-3& 0.0510E-3
%\end{tabular}
%\end{centering}
%\caption{Computed values of $\Delta_N$.}
%\label{tab:deltan}
%\end{table}

\subsection{Least squares fit, rate of decay}

\begin{figure}[ht] %  figure placement: here, top, bottom, or page
   \centering
(a)   \includegraphics[width=4.5in]{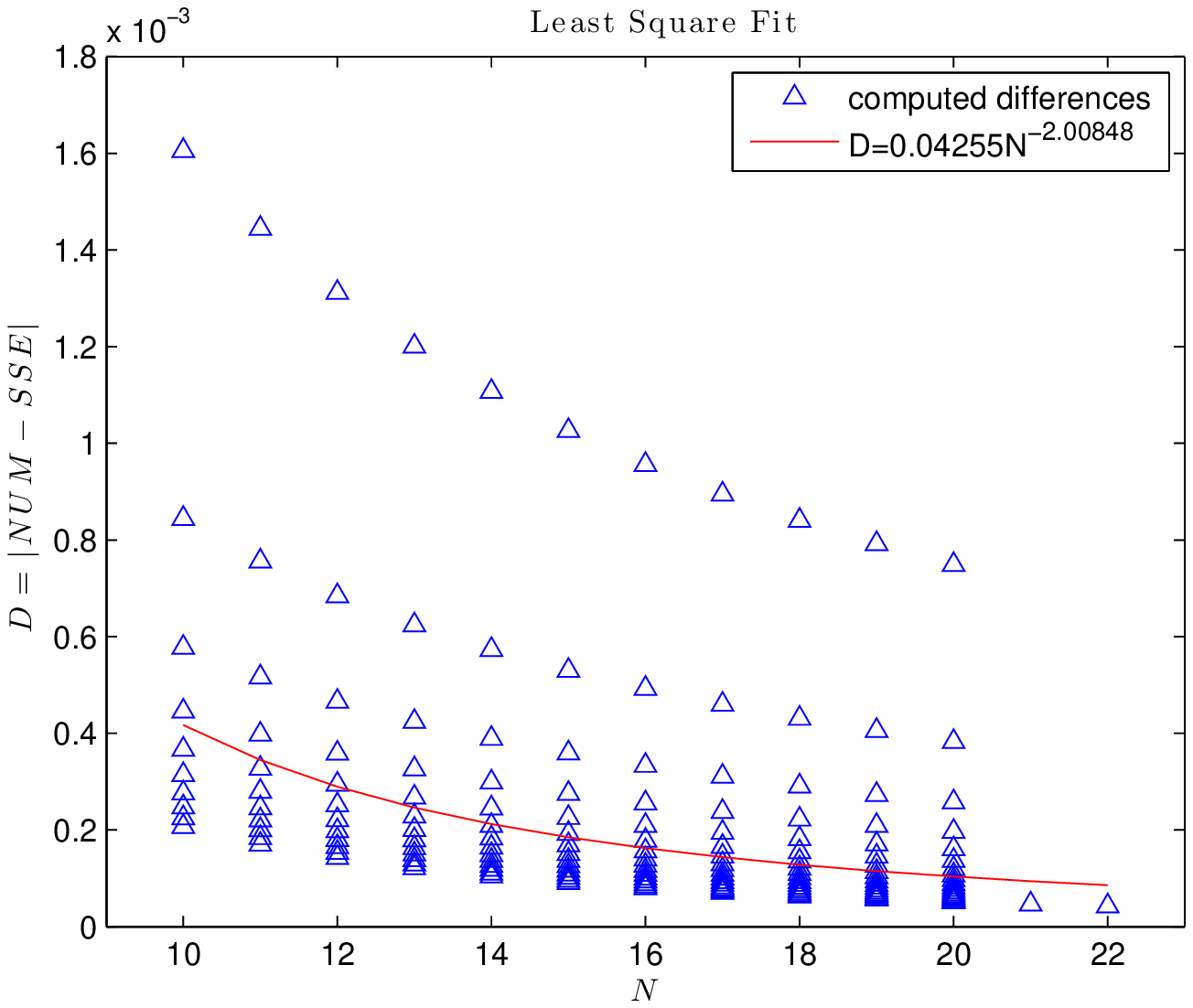} \\
(b)   \includegraphics[width=4.5in]{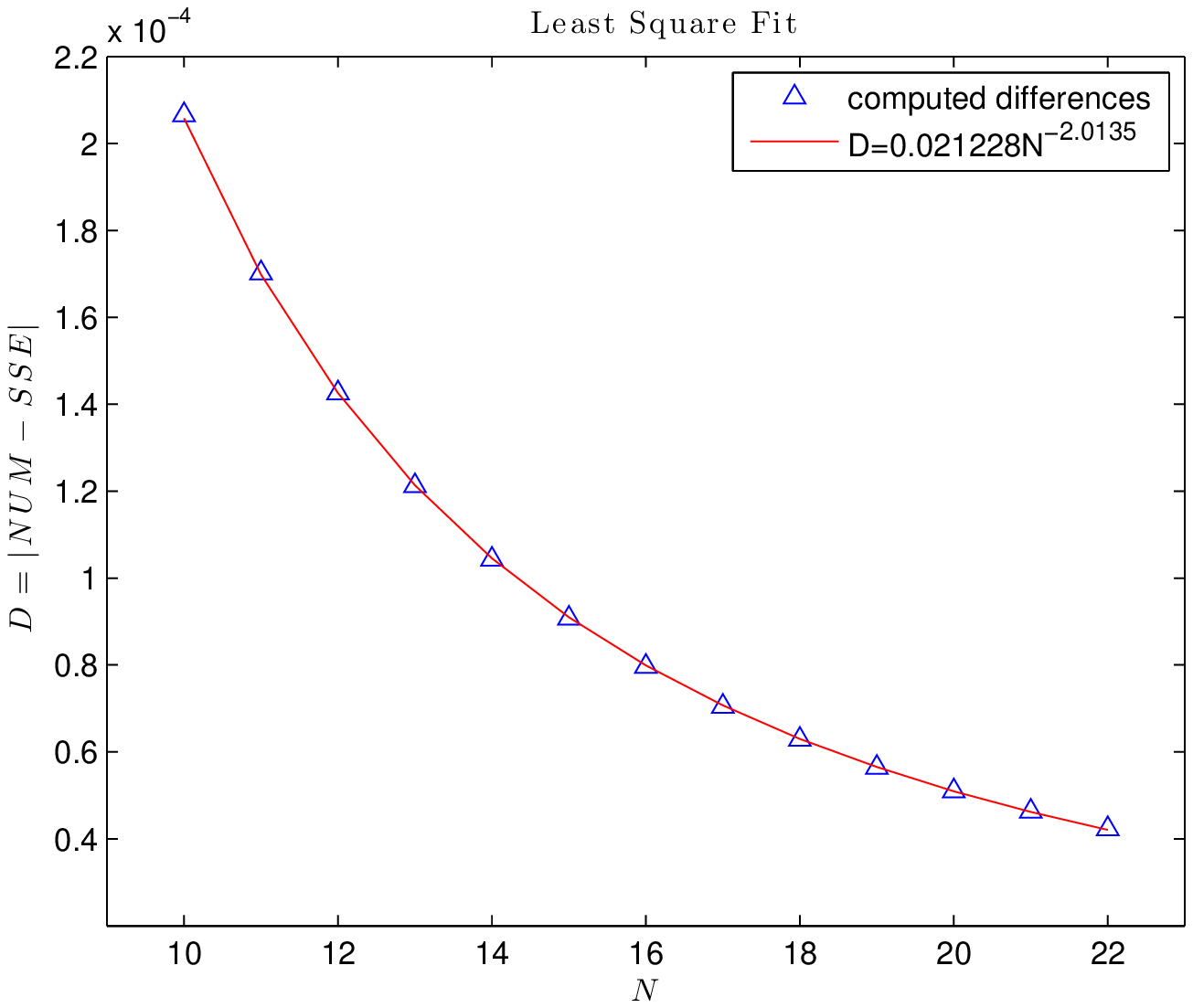} 
   \caption{(a) Least square fit for 167 data points, for which $N=10,\cdots, 20$ with $k=0,\ldots, N-1$, and $N=21, 22$ with $k=0$. (b) Least square fit for $k=0$, $N=10,\ldots, 22$.}
   \label{fig:least}
\end{figure}

We perform a least squares fit of the data in terms of
\beq\label{eq:D_least}
D_{*}^{N}=C_*\cdot N^{\alpha_*}\,,
\eeq
for some constants $C$ and $\alpha$. If we take the logarithmic function to the above equation, then the least squares fit is reduced to a linear least square fit in the $\log-\log$ space. Now for each $N=10,\cdots, 20$, we have data for $k=0,\cdots, N-1$, and for $N=21$ and 22, we have data for the largest ($k=0$) eigenvalue. Therefore we have total 167 data points. Figure \ref{fig:least}(a) shows the least square fit for all 167 data points. The triangles are the computed differences. For example, there are 10 eigenvalues for $N=10$, and hence there are 10 computed differences. The solid line is the  computed least square curve, which indicates the overall trend of decay of $D^{N}$ versus $N$. The rate of decay is $\alpha=-2.00848$.  We remark that the largest difference for each case of $N$ always occurs at the eigenvalue closest to the real axis ($k=N-1$), whereas the smallest difference occurs for the largest eigenvalue.
 
Another way to monitor the rate of decay for $D^{N}$ versus $N$ is to compute the difference of the largest eigenvalue for each $N$. That is, we compute 
\beq
D_{0}^{N} = |\lambda_{N,0}^\mathrm{app}-\lambda_{N,0}^\mathrm{wkb}|,\quad N=10,\cdots, 22.
\eeq 
Figure  \ref{fig:least}(b) is the least square fit for this collection of 13 data points. It shows that the rate of decay is $\alpha=-2.0135$. We are thus led to propose the following conjecture.

\begin{conjecture}\label{conj:eps2}
The WKB eigenvalues satisfy
\beq
|\lam_{N,k}^\mathrm{wkb}-\lam_{N,k}|=O(N^{-2})\quad\text{as}\;N\to\infty\,.
\eeq
Here, $\lambda_{N,k}^\mathrm{wkb}$ are the WKB eigenvalues given by \eqref{eq:bs} while $\lambda_{N,k}$ denote the true eigenvalues of \eqref{eq:zs} corresponding to the quantized values $\eps_N$. 
\end{conjecture}

This conjecture agrees with formal calculations of Miller \cite{M_PD01}; see the concluding discussion in \S\ref{ssec:prove}.

\section{Discussion}\label{sec:discuss}

\subsection{Future directions}

\begin{figure}[ht] %  figure placement: here, top, bottom, or page
   \centering
   \begin{tabular}{cc}
   (a)\includegraphics[width=2.8in]{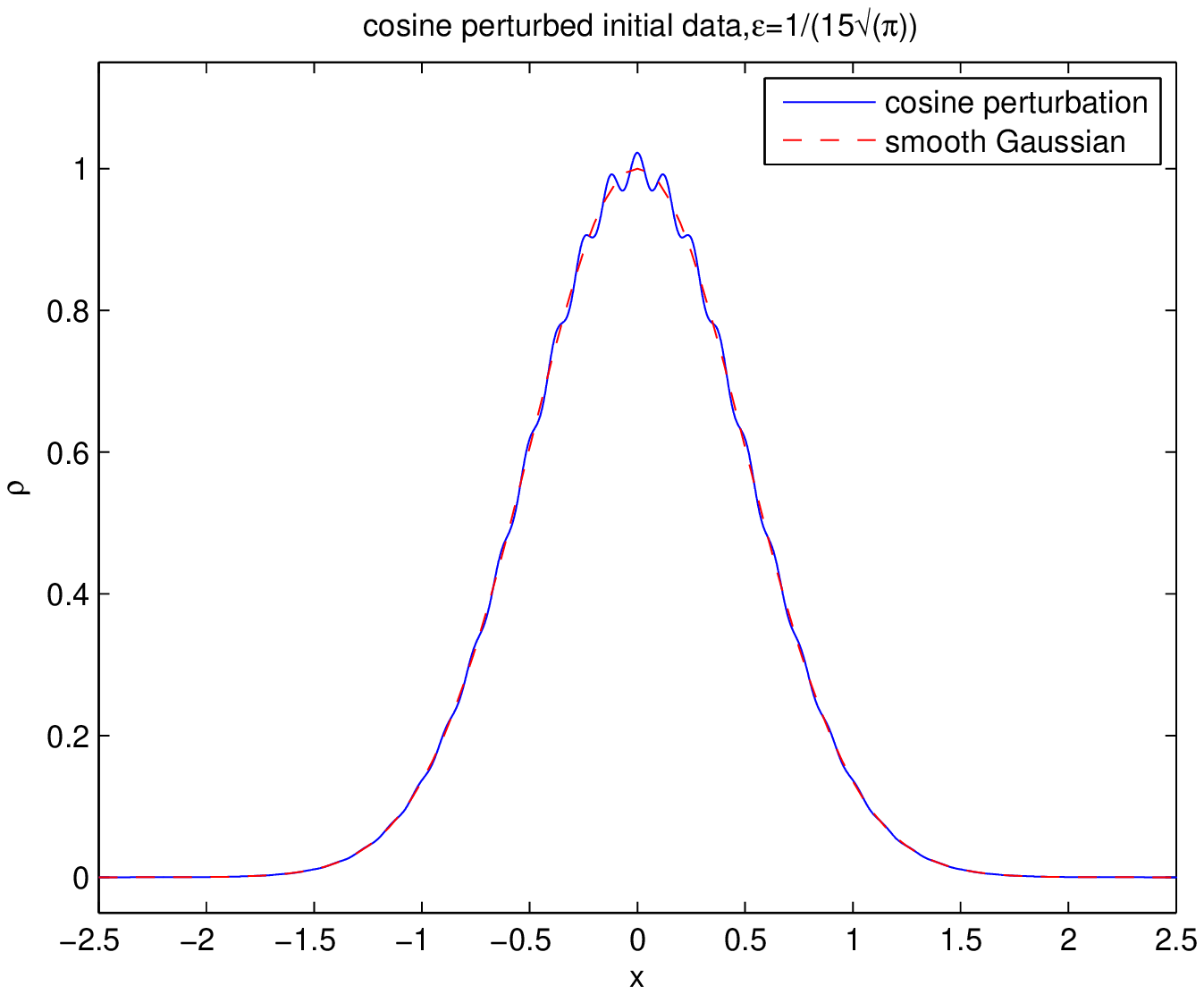} & 
   (b)\includegraphics[width=2.8in]{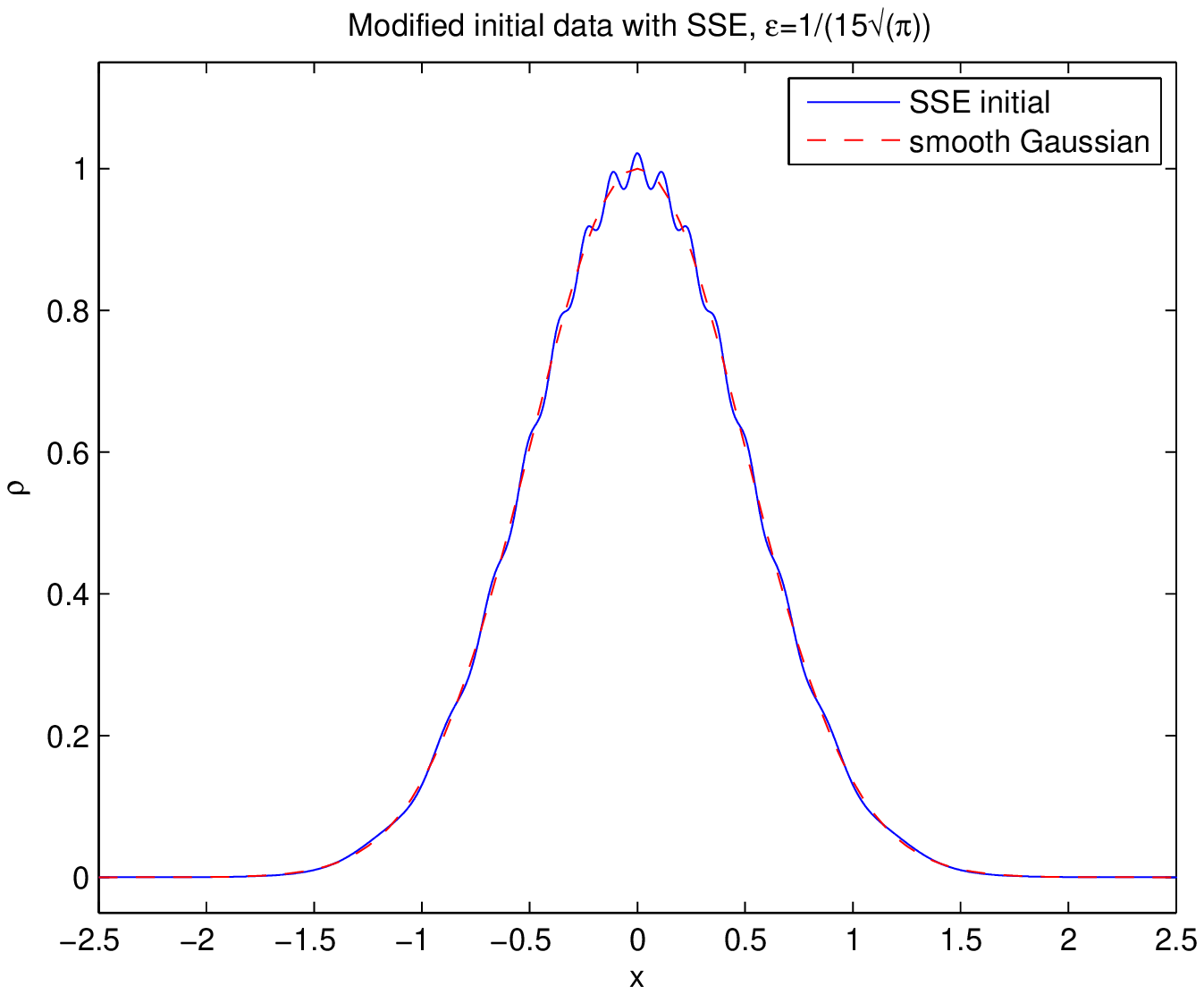}
   \end{tabular}
   \caption{(a) The cosine perturbation given in \eqref{eq:cosine}. (b) The reconstruction of the initial data $\psi_0^{(\eps)}$ using the WKB eigenvalues for $N=15$. Figures taken from \cite{LL_PLA13}.}
   \label{fig:cosine}
\end{figure}

One natural extension of this numerical experiment would be to try to use similar methods to examine the ``cosine-perturbed'' potentials used by Lee \& Lyng \cite{LL_PLA13} in their recent  study of the stability of the semiclassical limit. They considered potentials of the form 
\beq\label{eq:cosine}
\tilde{\psi}_0^{(\eps)}=0.3\cos\left(\frac{x}{0.54\eps}\right)\exp(-x^2)\,,\quad\eps>0\,;
\eeq
these were chosen to mimic the potentials $\psi_0^{(\eps)}$ that arise due to the use of the WKB eigenvalues. Figure \ref{fig:cosine} shows the close resemblance of of a member of the family of potentials in \eqref{eq:cosine} and the corresponding potential $\psi_0^{(\eps)}$. Lee \& Lyng found that, despite the superficial similarity between these two data, numerical simulations of the temporal evolution under the equation \eqref{eq:nls} appear to be extremely sensitive to the differences between the two potentials. That is, the differences appeared to almost instantaneously trigger the acute modulational instabilities known to be a feature of \eqref{eq:nls} in the small-$\eps$ regime. 
One possible explanation is that the spectrum of \eqref{eq:zs} is quite sensitive to the variations between perturbations of this kind. We observe (see Figure \ref{fig:cosine}) that the potentials in \eqref{eq:cosine} are \emph{not} single-lobe Klaus--Shaw potentials, and thus the spectrum need not be confined to the imaginary axis. Thus, we propose to revisit the spectral instability calculations of Bronski \cite{B}; his numerical results suggested that the eigenvalue problem with real potential is stable when subjected to nonanalytic perturbations. However, the focus there on analyticity is misleading; his nonanalytic perturbation was of Klaus--Shaw type. The cosine-perturbed potential provides an interesting example of an analytic but multiple-lobed potential.
%We have shown numerical evidence that the spectrum of the semiclassical ZS eigenvalue problem with an analytic potential and non-trivial phase is unstable to non-analytic perturbations. We have also shown that the spectrum of the ZS eigenvalue problem with a real potential appears to be stable to non-analytic perturbations.

In addition, this proposed numerical experiment provides an opportunity to develop and test the numerical techniques for Evans-function calculations aimed at detecting eigenvalues in exponentially asymptotic systems of the basic form 
\[
\frac{\dif}{\dif x}\vec{w}=\mat{A}(x;\lambda,\eps)\vec{w}\,.
\] 
Here, we have adopted the complex shooting method of Bronski \cite{B96}, but one intriguing possibility is to adopt some of the techniques from the Evans function community. A focus of this community has been on large systems (see, e.g., \cite{HSZ,HZ}), but preliminary work by Humpherys \& Lytle \cite{HL} on tracking eigenvalues by continuation is quite intriguing. Their continuation method would make it straightforward to follow eigenvalue branches as the parameter $\eps$ varies, and the oscillatory nature of the potential in \eqref{eq:cosine} provides a challenging test case for the developing numerical method. The results of this experiment might give some additional insight into the spectral origins of the modulational instability in \eqref{eq:nls}.

%\begin{enumerate}
%\item Cosine perturbation of \cite{LL_PLA13}, compare with Bronski's spectral stability of the problem \cite{B}. Non-Klaus--Shaw perturbation of the potential. 
%\item Incorporate Evans-function technology for improved numerical results. Humpherys \& Lytle \cite{HL} appear to have success with a ``continuation'' method. 
%\item Naturally, the next big step will be to prove the conjecture. (see below)
%\end{enumerate}

\subsection{Proving the conjecture and implications for the semiclassical limit }\label{ssec:prove}
%If the conjecture holds. First, one would like to actually prove it. (Peter's approach is based on a Langer transformation.) Recycle some ideas from \cite{M_PD01}, discuss the formal analysis from \cite{EJLM}
A second natural direction for future work would be to seek a rigorous proof of the $O(\eps^2)$ decay of the WKB eigenvalues to the true eigenvalues. Indeed, we believe that such a proof is highly likely to be an essential ingredient in the development of a complete theory for the semiclassical limit for \eqref{eq:nls} that is based on semiclassical soliton ensembles. Given that a completely rigorous theory is restricted to special, exactly solvable potentials, the extension to more general (bell-shaped, analytic) real data is a clearly worthwhile goal.  

Miller has given a possible roadmap for finding such a proof in the concluding discussion of his paper \cite{M_PD01}; in this paper he introduces a certain complexified WKB method for analyzing the spectrum of \eqref{eq:zs}. Although the analysis is formal, Miller's method is able to reproduce the $\mathsf{Y}$-shaped configurations of eigenvalues that Bronski \cite{B96} observed for potentials with a nontrivial phase $S_0$. As a starting point, Miller suggests a change of variables that transforms \eqref{eq:zs} to a Weber equation plus a small correction, and he speculates about the kind of tools from Kato's perturbation theory for linear operators \cite{Kato} that will be necessary to deal with the two-parameter family of linear operators that results from this plan of attack.
This program has not, to our knowledge, been carried out completely, but it seems to be a natural starting point. We believe that the new numerical evidence presented here provides additional impetus for pursuing this line of analysis. 

Finally, assuming the conjecture has been proved, an important next step will be to incorporate these error estimates into the asymptotic analysis of the semiclassical limit problem for \eqref{eq:nls}, as in \cite{KMM,LM}. However, we recall that a crucial step in this analysis is the ``sweeping away of the poles'' in a meromorphic Riemann--Hilbert problem (RHP). That is, one makes a change of variables which exchanges a meromorphic RHP for a sectionally holomorphic one. But, this change of variables is predicated on knowing the precise locations of the soliton eigenvalues. If the WKB approximations are used instead, this process will leave behind phantoms of the residues at these poles; the rate of decay in the conjecture provides a means of quantifying how quickly these phantoms disappear in the limit $\eps\downarrow0$.

\section*{Acknowledgement}
Research of YK and GL was supported in part by the National Science Foundation under grant number DMS-0845127.

\appendix
\section{GSSE and computed eigenvalues}\label{sec:wkbevals}

In this appendix, we report, in Tables~\ref{tab:N-10-11}--\ref{tab:N-20} below, the computed values of the locations of the eigenvalues given by the WKB formulae---the values of $\tau_{N,k}^\mathrm{wkb}$---and the corresponding values for $\tau_{N,k}^{\mathrm{app}}$ computed by Bronski's method described in \S\ref{sec:nm-Bronski} above for the case that
\[
\psi_0(x)=\me^{-x^2}\,.
\]
The values in these tables are exactly those used to create Figure \ref{fig:least}. For details of the computation of the WKB eigenvalues, see \cite{LLV}. 

\begin{table}[p]
\caption{WKB and Computed eigenvalues for $N= 10$ and $11$.}
\label{tab:N-10-11}
\begin{tabular}{c|cc|cc}
$k$ & $\tau_{10,k}^\mathrm{wkb}$  & $\tau_{10,k}^\mathrm{app}$ & $\tau_{11,k}^\mathrm{wkb}$ & $\tau_{11,k}^\mathrm{app}$ \\ \hline
0 &0.959902403980124800 &0.959695806284726 &0.963564788471487945 &0.963394619089388  \\ 
1 &0.878399870663193813 &0.878174978967203 &0.889623248317185619 &0.889439576040559  \\
2 &0.794927323219345668 &0.794679824110475 &0.814087692133196788 &0.813887684734036  \\
3 &0.709139284466368814 &0.708863047689429 &0.736712019197285355 &0.736491828432918 \\
4 &0.620568451131766213 &0.620254274771192 &0.657175161449747410 &0.656929289511540 \\
5 &0.528552832052961365 &0.528185937559299 &0.575043348681508226 &0.574763538606010 \\
6 &0.432092815727988847 &0.431647112406245 &0.489702969471305347 &0.489375947786600  \\ 
7 &0.329529022605841536 &0.328951524592522 &0.400229091055589894 &0.399831398024904  \\
8 &0.217634138896223524 &0.216790032806337 &0.305089744424815012 &0.304573671596604  \\  
9 &0.087541757627268806 &0.085936244035803 &0.201314538699416193 &0.200558457358652  \\ 
10 & & &0.080783014636351894 &0.079338649328913 
\end{tabular}
\end{table}

\begin{table}[p]
\caption{WKB and Computed eigenvalues for $N= 12$ and $13$.}
\label{tab:N-12-13}
\begin{tabular}{c|cc|cc}
$k$ & $\tau_{12,k}^\mathrm{wkb}$  & $\tau_{12,k}^\mathrm{app}$ & $\tau_{13,k}^\mathrm{wkb}$ & $\tau_{13,k}^\mathrm{app}$ \\ \hline
0 &0.966614106690141157 &0.966471514970761 &0.969192413823765906 &0.969071197539234 \\ 
1 &0.898948706858779235 &0.898795874675114 &0.906820384967780631 &0.906691231396956 \\
2 &0.829966905278888694 &0.829801882282211 &0.843342787387105468 &0.843204288407004 \\
3 &0.759487158306245807 &0.759307412149039 &0.778621850548315120 &0.778472277709511 \\
4 &0.687279237207741584 &0.687081282016942 &0.712486801434484228 &0.712323839605675 \\
5 &0.613043064498824122 &0.612821921069498 &0.644721236776553608 &0.644541703425410 \\
6 &0.536373677360965851 &0.536121862724695 &0.575043348681508226 &0.574842696621590 \\
7 &0.456698971984012495 &0.456404446862373 &0.503073222683235512 &0.502844613911967\\
8 &0.373157976887155824 &0.372799442399859 &0.428274801093616700 &0.428007225964664\\
9 &0.284327593774234001 &0.283861684652851 &0.349842567456668418 &0.349516532372660 \\
10 &0.187456221347599428 &0.186772191716481 &0.266447608380459765 &0.266023383069821 \\
11 &0.075054860278860741 &0.073743084165890 &0.175527000002819191 &0.174902970556353 \\
12 & & &0.070133331675973478 &0.068932484184659
\end{tabular}
\end{table}

\begin{table}[p]
\caption{WKB and Computed eigenvalues for $N= 14$ and $15$.}
\label{tab:N-14-15}
\begin{tabular}{c|cc|cc}
$k$ & $\tau_{14,k}^\mathrm{wkb}$  & $\tau_{14,k}^\mathrm{app}$ & $\tau_{15,k}^\mathrm{wkb}$ & $\tau_{15,k}^\mathrm{app}$ \\ \hline
0 &0.971401021947088984 &0.971296712115742 &0.973314130299657922 &0.973223400764770\\
1 &0.913553804153339020 &0.913443229185553 &0.919379303197748057 &0.919283577415958 \\
2 &0.854764915853138965 &0.854647010795885 &0.864632747557207840 &0.864531155636158 \\
3 &0.794927323219345668 &0.794800875238435 &0.808989649893436479 &0.808881324093153 \\
4 &0.733910823730560734 &0.733774238016687 &0.752348450611801224 &0.752232260635562 \\
5 &0.671554128233757039 &0.671405275616136 &0.694585672871886988 &0.694460141475419 \\
6 &0.607652999485902664 &0.607488954095893 &0.635548451445279806 &0.635411609391642 \\
7 &0.541941693882958848 &0.541758273539448 &0.575043348681508226 &0.574892489650501 \\
8 &0.474062300679528845 &0.473853212686411 &0.512818870241486763 &0.512650124248260 \\
9 &0.403510388842961381 &0.403265495038545 &0.448536604967941901 &0.448344146100832 \\
10 &0.329529022605841536 &0.329230359378978 &0.381720116930106813 &0.381494553787561 \\
11 &0.250871627712129448 &0.250482550416875 &0.311655409515556435 &0.311380091074583 \\
12 &0.165139740648526037 &0.164566405034957 &0.237168544019441450 &0.236809476680163 \\
13 &0.065855659394471488 &0.064748933697033 &0.156005735208995840 &0.155475763883866 \\
14 & & &0.062100574201615084 &0.061074675232078
\end{tabular}
\end{table}

\begin{table}[p]
\caption{WKB and Computed eigenvalues for $N= 16$ and $17$.}
\label{tab:N-16-17}
\begin{tabular}{c|cc|cc}
$k$ & $\tau_{16,k}^\mathrm{wkb}$  & $\tau_{16,k}^\mathrm{app}$ & $\tau_{17,k}^\mathrm{wkb}$ & $\tau_{17,k}^\mathrm{app}$ \\ \hline
0 &0.974987326652522948 &0.974907679543992 &0.976463078563721420 &0.976392582892975 \\
1 &0.924468977718866355 &0.924385312070535 &0.928954000679293301 &0.928880268976254 \\
2 &0.873243665753901065 &0.873155219305772 &0.880823661705302338 &0.880745966869343 \\
3 &0.821243041488997917 &0.821149186513302 &0.832016171949275511 &0.831934059201085 \\
4 &0.768386349584278361 &0.768286267983453 &0.782466254532225472 &0.782379123773133 \\
5 &0.714576934358073508 &0.714469571007124 &0.732096814876746890 &0.732003895226637 \\
6 &0.659697341348599604 &0.659581322192200 &0.680815619395365766 &0.680715923722662 \\
7 &0.603602244506645897 &0.603475738765136 &0.628510641973456798 &0.628402886756366 \\
8 &0.546107865626943024 &0.545968356699761 &0.575043348681508226 &0.574925823934691 \\
9 &0.486975445955067848 &0.486819333738103 &0.520238658424880569 &0.520109012983977 \\
10 &0.425883986962532764 &0.425705849388591 &0.463869271083202648 &0.463724140310799 \\
11 &0.362382024028150845 &0.362173116393630 &0.405629841382795024 &0.405464155846148 \\
12 &0.295793808177383880 &0.295538617435822 &0.345091329157721534 &0.344896909373886 \\
13 &0.225009980976628986 &0.224676814993881 &0.281612286351431294 &0.281374614927273 \\
14 &0.147905057406870792 &0.147412579070145 &0.214141169841336466 &0.213830570790340 \\
15 &0.058775789787028711 &0.057820024496162 &0.140667075044501901 &0.140207316723819 \\
16 & & &0.055809776952142479 &0.054915420231151
\end{tabular}
\end{table}

\begin{table}[p]
\caption{WKB and Computed eigenvalues for $N= 18$ and $19$.}
\label{tab:N-18-19}
\begin{tabular}{c|cc|cc}
$k$ & $\tau_{18,k}^\mathrm{wkb}$  & $\tau_{18,k}^\mathrm{app}$ & $\tau_{19,k}^\mathrm{wkb}$ & $\tau_{19,k}^\mathrm{app}$ \\ \hline
0 &0.977774388702513135 &0.977711530027987 &0.978947292585139432 &0.978890864294410\\
1 &0.932936107055917977 &0.932870661496479 &0.936495415330534105 &0.936436960601441 \\
2 &0.887547565589977046 &0.887478780329797 &0.893552760640799980 &0.893491445327538 \\
3 &0.841562477890611001 &0.841490026938481 &0.850080563544862558 &0.850016160482089 \\
4 &0.794927323219345668 &0.794850764431408 &0.806034393653292551 &0.805966579478932 \\
5 &0.747579613280034652 &0.747498374113044 &0.761362892085452425 &0.761291237429385 \\
6 &0.699445575060456649 &0.699358929568178 &0.716006114657257995 &0.715930075899816 \\
7 &0.650436985872799092 &0.650344007644111 &0.669893317782499000 &0.669812210227375 \\
8 &0.600446741113353703 &0.600346226387779 &0.622939935595656631 &0.622852886952360 \\
9 &0.549342461771020473 &0.549232806734786 &0.575043348681508226 &0.574949225838631 \\
10 &0.496956943375841810 &0.496835942283868 &0.526076784575294106 &0.525974077672774 \\
11 &0.443073256283906445 &0.442937751607330 &0.475880209617196262 &0.475766841639301 \\
12 &0.387400210236114836 &0.387245442960016 &0.424246129054545068 &0.424119125938343 \\
13 &0.329529022605841536 &0.329347311254836 &0.370896220337988608 &0.370751098277158\\
14 &0.268849192277737029 &0.268626897603966 &0.315440098159880707 &0.315269617599447\\
15 &0.204361251874907954 &0.204070478194082 &0.257295332382745482 &0.257086634092119\\
16 &0.134157265068993706 &0.133726296173683 &0.195509617551811908 &0.195236389001754\\
17 &0.053146200608543349 &0.052306042599784 &0.128268066850698152 &0.127862613252463\\
18 & & &0.050740062467130075 &0.049948078507488
\end{tabular}
\end{table}

\begin{table}[p]
\caption{WKB and Computed eigenvalues for $N= 20$, 21, and 22.}
\begin{tabular}{c|cc|cc}
$k$ & $\tau_{20,k}^\mathrm{wkb}$  & $\tau_{20,k}^\mathrm{app}$ &$\tau_{21,k}^\mathrm{wkb}$  & $\tau_{21,k}^\mathrm{app}$ \\ \hline
0 &0.980002604979491981 &0.979951631457158 &0.980957166067443195 &0.980910847289176 \\
1 &0.939695880480300898 &0.939643385837989 & & \\
2 &0.898948706858779235 &0.898893721170934 & & \\
3 &0.857728294094450955 &0.857670668045702 & & \\
4 &0.815997352062456992 &0.815936855083385 & & \\
5 &0.773713152397110641 &0.773649462044337 & & \\
6 &0.730826318228443458 &0.730759022210825 & & \\
7 &0.687279237207741584 &0.687207819788240 & & \\
8 &0.643003941549260073 &0.642927755401035 & & \\
9 &0.597919214876281576 &0.597837436051694 & & \\
10 &0.551926544341512482 &0.551838102644521 & & \\
11 &0.504904288332439020 &0.504807757551170 & & \\
12 &0.456698971984012495 &0.456592389770675 & & \\
13 &0.407111724608396028 &0.406992280447767 & & \\
14 &0.355875976460279444 &0.355739432084045 & & \\
15 &0.302618129825466500 &0.302457640314857 & & \\
16 &0.246781339625886001 &0.246584742991723 & & \\
17 &0.187456221347599428 &0.187198621768910 & & \\
18 &0.122912395848409971 &0.122529702271993 & &  \\ 
19 &0.048554967111164647 &0.047806072582520 &  & \\ \hline
& & &$\tau_{22,k}^\mathrm{wkb}$  & $\tau_{22,k}^\mathrm{app}$\\ \hline
0 & & &0.981824746922388093 &0.981782417756205
\label{tab:N-20}
\end{tabular}
\end{table}
\bibliographystyle{plain}
\bibliography{llv}

\end{document}